\documentclass[aps,prl,reprint,showpacs]{revtex4-1}

\usepackage{color}
\usepackage{graphicx}
\usepackage{amsmath}

\begin{document}
\title{Almost local generation of EPR entanglement}
\date{\today}
\author{Rebecca Schmidt}
\email{rebecca.schmidt@uni-ulm.de}
\affiliation{Institut f\"{u}r Theoretische Physik, Universit\"{a}t Ulm,
  Albert Einstein-Allee 11, 89069 Ulm,
  Germany}
\author{J\"urgen T. Stockburger}
\affiliation{Institut f\"{u}r Theoretische Physik, Universit\"{a}t Ulm,
  Albert Einstein-Allee 11, 89069 Ulm,
  Germany}
\author{Joachim Ankerhold}
\email{Current address: Department of Physics, Dartmouth College, Hanover (NH), USA.}
\affiliation{Institut f\"{u}r Theoretische Physik, Universit\"{a}t Ulm,
  Albert Einstein-Allee 11, 89069 Ulm,
  Germany}

\begin{abstract}
  The generation of entanglement is studied in a minimal model
  consisting of two independent Gaussian parties embedded in a common
  heat bath. We consider the case of weak reservoir-induced
  interactions which themselves are insufficient to generate
  entanglement. Local driving by an external classical field, however,
  can promote this weak interaction to a source of
  entanglement. Presence or absence of the effect depends on the
  specific pulse shape of the external control, which we determine
  through optimal control techniques.
\end{abstract}
\pacs{03.65.Yz, 03.67.Bg, 02.30.Yy}

\maketitle
\paragraph{Introduction.-}Entanglement is one of the most fascinating
manifestations of quantum non-locality in systems with more than one
degree of freedom. It has thus been investigated as a key resource for
quantum information in case of both discrete
\cite{bennett1993,
NiCh2000
} and continuous variables
\cite{Vaidmann1994
,Braunstein2005
,Weed2012
,DAur2009,
Laur2005%
}.
Both fundamental properties of bipartite Gaussian entanglement and
dynamical process producing entangled states have been investigated
extensively%
\cite{Ades2004
  ,paz08
  ,Eis2002
  ,Oliv2011
  ,Hoer2008
  ,Wolf2011
}, with varying roles assigned to a dissipative environment, and to
external control fields~%
\cite{Galve2010
  ,McEn2012
  ,Sera2010
  ,Zipp2013
}.

The non-local interaction required for the generation of
entanglement is sometimes postulated as an explicit feature of the
system~%
\cite{paz08
  ,Sera2009
  ,Galve2010
}. Alternatively, entanglement can be generated through forcing the
system state towards the pointer states of a strongly dissipative
reservoir, or by relying on the effective interaction potential
arising from the exchange of virtual reservoir excitations~%
\cite{McEn2012
  ,Wolf2011
}. The latter type of interaction may be strong even in the case of
weak dissipation; its main features can be revealed through the
adiabatic elimination of high-frequency reservoir excitations.
The cases of strong dissipation or strong environment-induced coupling
may be undesirable or infeasible in intended applications. The
entangled states thus produced are useful only if the strong
interactions that produced them can reliably be quenched, i.e., if the
system-reservoir coupling can be controlled at will.

In this Letter, we therefore consider a more subtle dissipative
mechanism which will not generate entanglement by itself. However,
when leveraged by suitably chosen classical driving fields, the
combined effect of driving and dissipation can result in entanglement.
We do not assume environments which require specific engineering \cite{Arenz2013},
but rather are known as standard models for decoherence in solid state devices \cite{Makhlin2001,Weiss2008}.
Remarkably, even \emph{local} parametric driving is sufficient to
promote the reservoir from an entanglement-degrading to an
entanglement-promoting feature. This effect is quite sensitive to the
exact time dependence of the driving fields, which we determine
through optimal control techniques~\cite{epaps,RS2011}.

\begin{figure}[b]
\includegraphics[width=0.67\linewidth]{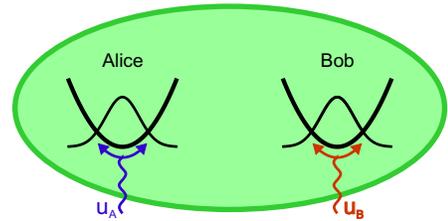}
\caption{\label{fig:logo} Two independent harmonic modes A and B are
  embedded in a common reservoir (green), which equilibrates the modes
  to a {\em non-entangled} stationary state. Time-dependent control fields
  applied \emph{locally} to the A and B change the reservoir into a
  source of entanglement.}
\end{figure}

We find that a weakly damped bipartite harmonic system in a Gaussian
state (Fig.~\ref{fig:logo}) can be driven into an entangled state by
local driving. The immediate effect of local driving on modes
A and B is single-mode squeezing. Dissipation in the reduced dynamics
of the bipartite system transforms this single-mode squeezing into
two-mode squeezing (when viewing symmetric and antisymmetric
modes) or entanglement (when viewing local modes A and B). This
effect is found over a wide temperature range. Suitable pulse shapes
yielding significant entanglement can be found even for the case of
driving only mode A.  These findings are of potential relevance for
current experiments with superconducting circuits. They also open new possibilities for teleportation without an explicit state transfer.

\paragraph{Open system dynamics.-}
We consider a compound system with Hamiltonian $H=H_S(t)+H_{I}+H_{R}$,
where two harmonic oscillators of equal mass $M$ and frequency
$\Omega$ form a distinguished system with
\begin{equation}\label{eq:systemH}
H_S(t)\equiv H_{{A}}(t)+H_{{B}}(t)=\sum_{j=A,B} \frac{p_{j}^2}{2
M}+\frac{M \Omega^{2}}{2}q_{j}^{2}+\frac{u_{j}(t)}{2}q_{j}^{2}.
\end{equation}
The last term in Eq. (\ref{eq:systemH}) represents local parametric
driving of strength $u_j(t)$. The oscillators interact with a common
reservoir which has the conventional form of a thermal reservoir used
in a quantum Brownian motion context
\cite{CaLe1983,Weiss2008,Hoer2008}, i.e.,
\begin{eqnarray}
H_{R}&=&\sum_{k} \frac{p_{k}^2}{2
  m_{k}}+\frac{m_{k}\omega_{k}^{2}}{2}x_{k}^{2}\label{eq:HR}\\
  H_{I}&=&(q_{A}+q_{B})\sum_{k}c_{k}x_{k}+(q_{A}+q_{B})^{2}\sum_{k}
  \frac{c_{k}^2}{2 m_{k}\omega_{k}^{2}}\, .\label{eq:HI}
\end{eqnarray}
The reduced dynamics depends on the properties of the reservoir only
through the inverse thermal energy $\beta$ and through the spectral
density $J(\omega)= \pi \sum_k c_k^2/(2 m_k \omega_k)
\delta(\omega-\omega_k)$, formed from the parameters in
Eqs. (\ref{eq:HR}) and (\ref{eq:HI}).  The last part in $H_I$ (known
as ``counterterm'' in the context of the Caldeira-Leggett model)
should not be misread as an interaction term for the modes; it ensures
a \emph{vanishing} net effect of the reservoir on the dynamics if
adiabatic elimination is performed. If the reservoir is traced out
from the exact, full dynamics, only reservoir fluctuations and
velocity-dependent memory friction affect the system dynamics. In the
present two-mode model, the memory friction includes ``mutual drag''
between the oscillators, which is the only feature of the model that
qualifies as a source of quantum non-locality~\cite{Hoer2008}.

While the complete density operator of system and reservoir obeys the
standard Liouville-von Neumann equation, the treatment of the dynamics
of the relevant reduced density is a formidable challenge. This is
particularly true in case of low temperatures and {\em a priori}
unknown control signals where conventional perturbative expansions
like Born-Markov master equations fail. In this situation stochastic
Liouville von-Neumann equations (SLN) have been proven as formally
exact and numerically powerful tools to capture on the same footing
the non-Markovianity of the reduced time evolution and arbitrary
external driving fields also for nonlinear systems. They are based on
a stochastic representation of the Feynman-Vernon influence functional
such that the physical reduced density is obtained by averaging the
time evolution according to the SLN over proper noise
realizations~\cite{Stock2002}.

For Gaussian modes, the reduced density matrix is fully determined by
 the first and second cumulants of the elements $x_j, j=1, \ldots 4$
 of the vector $x=(q_A, p_A, q_B, p_B)$ of phase space
 operators. Since first moments can always be adjusted by local
 unitary operations, they cannot affect entanglement properties. The
 central quantity is thus the covariance matrix $\sigma$ with
 $\sigma_{ij}=\frac{1}{2}\langle x_{i}x_{j}+x_{j}x_{i}\rangle-\langle
 x_{i}\rangle\langle x_{j}\rangle$. In the
$2\times 2$ block structure~\cite{Ades2004}
\begin{eqnarray}
\sigma=\begin{pmatrix}
\alpha & \gamma\\
\gamma^{T} & \beta
\end{pmatrix}\, ,
\end{eqnarray}
$\alpha$, $\beta$ correspond to the covariance matrices of the
respective sub-units A and B, while $\gamma$ (transpose $\gamma^T$)
carries the mixed cumulants and thus non-local information.  Now, a
well-defined measure for entanglement in bipartite Gaussian systems is
given by the logarithmic negativity~\cite{Vida2002, Ades2004}
\begin{equation}
E_{\mathcal{N}}=\max \{0,-\ln(\tilde{\nu}_-)\},
\end{equation}
with $\tilde{\nu}_-$ being the smallest symplectic eigenvalue of the
partially transposed density matrix. Entanglement only exists if
$E_{\mathcal{N}} > 0$ while $E_{\mathcal{N}} = 0$ corresponds to
purely classical and/or local quantum correlations.  Note that $\det
\gamma < 0$ is necessary for $E_{\mathcal{N}}$ to be
positive~\cite{Sim2000}.

\begin{figure}
\includegraphics[width=\linewidth]{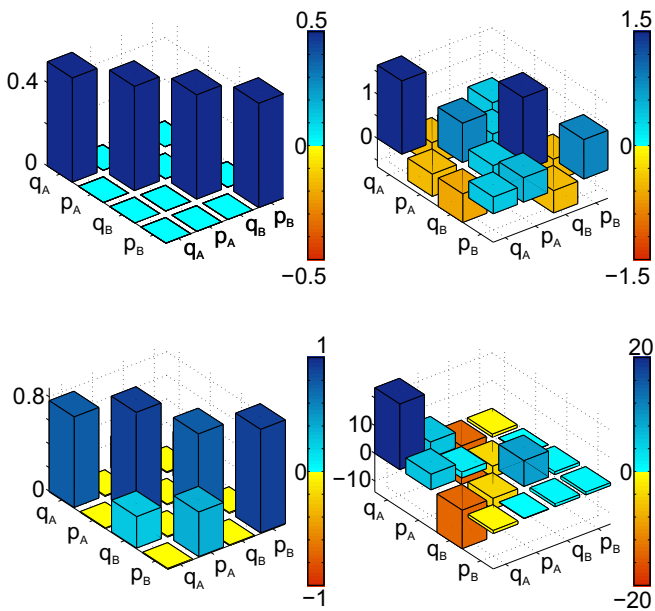}
\caption{\label{fig:CM} Covariance matrices for reservoir
  parameters $\beta =1$, $\eta=0.1$, and $\omega_c=50$. Left:
  autonomous dynamics; ground state as initial state (top), stationary
  final state (bottom). Right: final states of parametrically driven
  dynamics with $u_{A}=u_{B}$ (top) and $u_{A}\ne 0$, $u_{B}=0$
  (bottom) at $t_f=6\pi$. Dimensionless units are used, see text for
  details.}
\end{figure}

Without external driving, any initially separable two-mode state will
remain separable indefinitely in the scenario considered here. This
applies in particular to the steady state, which can be calculated
non-perturbatively exactly~\cite{Weiss2008}. The resulting expression
$\tilde{\nu}_-^2=\langle (q_A+q_B)^2\rangle_\beta \langle
(p_A+p_B)^2\rangle_\beta/(\hbar/2)^2$ contains an uncertainty product
of the thermal variances {\em always} exceeding the ground state limit
(cf.~Fig.~\ref{fig:CM}) so that $\det\gamma>0$. It is to be noted,
however, that this does not preclude discord as a further type of
non-classical correlations~\cite{Ades2010, Gior2010}.

\paragraph{Optimal control of entanglement.-} Although we have stated
that interaction with a common reservoir alone is not sufficient to
generate entangled states, it nevertheless plays a key role in the
generation of entanglement through an external control signal: Since
the system Hamiltonian (\ref{eq:systemH}) provides no two-body
interactions (neither static nor controlled), it is needed as an
additional ingredient beyond local controls. In the sequel, we not
only show that entangled states can be created by this combination of
factors, but also perform a numerical search for the maximal
entanglement that can be achieved in finite time. For this purpose,
the recently developed optimal control formalism for open quantum
systems~\cite{RS2011} is exploited to maximize the entanglement at a
given final time $t_f$, i.e., we seek a maximum of the functional
\begin{equation}
F[u_A(t), u_B(t);\sigma(t)]=E_{\mathcal{N}}(\sigma (t_f))
\end{equation}
of the control fields $u_A(t), u_B(t)$.

We consider an initial preparation with both oscillators in the ground
state, and without initial correlations between system and
reservoir. This is followed by propagation of the system under the
influence of control fields $u_A(t)$, $u_B(t)$ and interaction with an
Ohmic bath, $J(\omega)=\frac{\eta
\omega}{(1+(\omega^{2}/\omega_c^{2}))^{2}} $, parameterized by a
coupling constant $\eta$, high frequency cut-off $\omega_c$ and at
inverse temperature $\beta=1/k_{\rm B} T$.  We use dimensionless units
with frequencies scaled with $\Omega$ and lengths scaled with
$\sqrt{\hbar/M\Omega}$.

Fig.~\ref{fig:CM} displays bar graphs of
optimized covariance matrices (right) compared to ground and stationary
states (left). Apparently, substantial off-diagonal
correlations are built up, positive and negative valued, so that one
indeed has $\det\gamma<0$. Notably, the same is true for single-site
control ($u_B(t)\equiv 0$).

\begin{figure}
\includegraphics[width=1\linewidth]{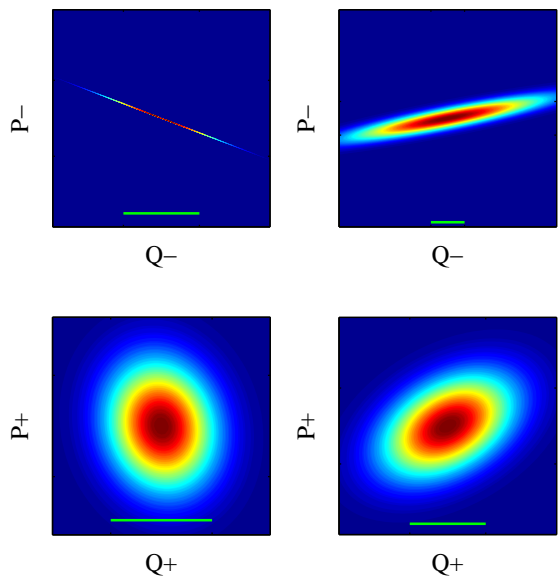}
\caption{\label{fig:wigner} Reduced Wigner functions for the
anti-symmetric $Q_-, P_-$ (top) and symmetric $Q_+, P_+$ (bottom)
normal modes at $t_f=6\pi$ for two-site (left) and single-site (right)
control. The green line indicates the ground state width; parameters
are as in Fig.~\ref{fig:CM}.}
\end{figure}

Further insight is gained through a transformation to normal modes
$Q_\pm=(q_A\pm q_B)/\sqrt{2}$, $P_\pm=(p_A\pm p_B)/\sqrt{2}$ which
transforms the system Hamiltonian (\ref{eq:systemH}) into
$\tilde{H}_S=\sum_{\alpha=\pm} \frac{P_\alpha^2}{2}
+\frac{1}{2}(1+\frac{u_A+u_B}{2})Q_\alpha^2 + \frac{1}{2}(u_A-u_B) Q_+
Q_-$. Then, for $u_A(t)\equiv u_B(t)$, entanglement
($E_{\mathcal{N}}>0$) requires
\begin{equation}\label{eq:detgam}
4\det\gamma =\Delta_P\, \Delta_Q -[\langle Q_+ P_+\rangle-\langle Q_- P_-\rangle]^2
\end{equation}
to be negative, where $\Delta_X=\langle X_+^2\rangle-\langle
X_-^2\rangle, X=P, Q$. Apparently, this is only possible if the
evolution of the two normal modes is not degenerate. Lifting this
degeneracy is the decisive role the heat bath plays in this
setting. However, external driving is needed as well in order
to render the normal mode states dissimilar enough to result in
entangled A and B states.

Entanglement generation is thus a \emph{cooperative effect} of local
driving and global dissipation in the present setting---either factor
by itself is neutral or detrimental to entanglement. A tailored
control pulse changes the non-local effects of the heat bath from
a destructive influence on quantum resources to an asset promoting
entanglement.

The correlators in Eq. (\ref{eq:detgam}) can be obtained from the
Wigner functions plotted in Fig.~\ref{fig:wigner} for the final state
under an optimized driving protocol. For symmetric control (left),
$u_A \equiv u_B$, a strongly squeezed antisymmetric mode results,
while the symmetric mode is close to a thermal state. This leads to
opposite signs of the terms $\Delta_P$ and $\Delta_Q$, therefore the
r.h.s. of Eq. (\ref{eq:detgam}) is negative, and the local modes $q_A$
and $q_B$ are entangled. For a more quantitative
analysis~\cite{epaps}, it is convenient to represent the correlation
matrix elements $\langle Q^2_{\pm}\rangle$, $\langle P^2_{\pm}\rangle$
and $\langle Q_{\pm}P_{\pm}\rangle$ by a different parameter set:
squeezing parameters $r_{\pm}$, squeezing angles $\varphi_{\pm}$, and
width parameters $a_{\pm}$. Using these parameters, one finds that
$\det \gamma< 0$ if the difference $|r_-- r_+|$ is sufficiently large,
with a threshold that depends on $\varphi_- - \varphi_+$ and on the
ratio $a_-/a_+$. This observation shows a similarity between the
states considered here and two-mode EPR states: EPR states can be seen
as factorized pure states of squeezed symmetric and antisymmetric
modes, with $r_- = - r_+$ and identical squeezing
angles~\cite{Leonhardt2010}. The states produced in our numerical
simulations show strong squeezing in the antisymmetric mode, but
\emph{not} in the symmetric mode, which is close to a thermal
state. They might therefore be labeled `semi-EPR' states.

Even using a more rudimentary \emph{single-site} control ($u_A\neq
0, u_B\equiv 0$), we find that the cooperative effect between driving
and dissipation persists, although the similarity to EPR states is
diminished, see Fig.~\ref{fig:wigner} (right). Moreover, additional
correlations between the symmetric and antisymmetric modes are needed
to fully characterize the quantum state.
\begin{figure}
\includegraphics[width=0.9\linewidth]{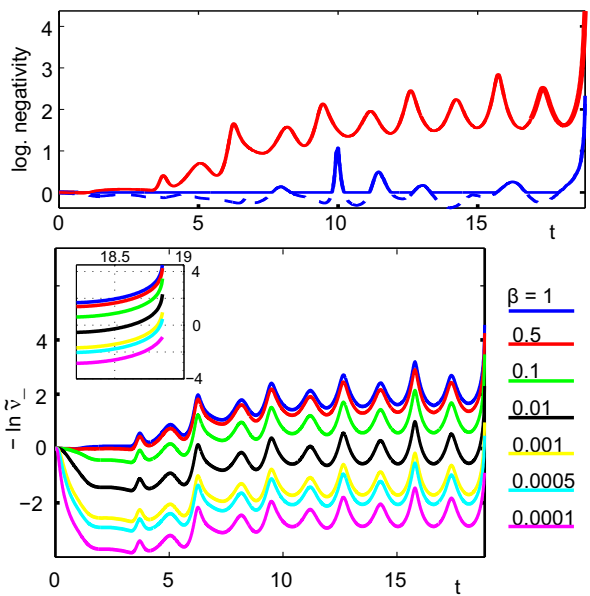}
\caption{\label{fig:Ln}Top: Logarithmic negativity $E_{\mathcal{N}}(t)$ (solid) for two-site (red) and single-site (blue) control.
Negative values for  $-\ln (\tilde{\nu}_-)$ corresponding to
$E_{\mathcal{N}}(t)=0$ are also depicted (dashed). Parameters are as in
Fig.~\ref{fig:CM}. Bottom:  $-\ln (\tilde{\nu}_-)$ for various temperatures
(two-site control)
with other parameters held fixed. The inset displays the final time range for the control with \ $E_{\mathcal{N}}(t_f)>0$  for $\beta\geq 0.0005$.}
\end{figure}
The gradual (non-monotonous) build-up of entanglement over time is
shown for the cases of symmetric and single-site control in
Fig.~\ref{fig:Ln} (top), with a rapid final approach to values of
$E_{\mathcal{N}}(t_f)=2.33$ (single-site control) and
$E_{\mathcal{N}}(t_f)=4.37$ (symmetric control).
Simulations at lower temperatures show further improvements.

The numerical values of the entanglement measure may be related to the
number of states involved, allowing a rough comparison to qubit-based
entanglement. For two qubits, $E_{\mathcal{N}}=1$ corresponds to a
Bell state. Furthermore, for two-mode squeezed vacuum states with
squeezing parameter $r$ and negativity $E_{\mathcal{N}}(r)=2 r$, the
number of excited states in each mode dominantly contributing to the
entanglement can be estimated as
$m\approx\exp[E_{\mathcal{N}}(r)]/2$. While, as discussed above, the
situation here is different due to the dissipative $Q_+$ mode, one may
at least estimate that symmetric (single-site) control involves about
$m\approx 40$ ($m\approx 5$) states in each of the oscillators.

Let us now discuss the temperature dependence of the
protocol. Qualitatively, one may expect quantum non-locality to be
quite robust, particularly for $u_A=u_B$. This is indeed seen in
Fig.~\ref{fig:Ln} (bottom) where the logarithmic negativity is of
order 1 even for inverse temperatures around $\beta=0.01$. Further
understanding can be gained by observing that the fundamental
solutions of the classical equation of motion play a prominent role
even in the fluctuations of a harmonic
oscillator~\cite{Weiss2008}. Generalizing results for the
parametrically driven oscillator~\cite{zerbe1995}, a semi-analytical
theory for the case of symmetric driving $u(t) = u_A(t)=u_B(t)$ can be
found based on the fundamental solutions $\phi_{1,2}(t)$ of the
equation of motion $\ddot{\phi}+[1+u(t)] \phi=0$.

\begin{figure}
\includegraphics[width=\linewidth]{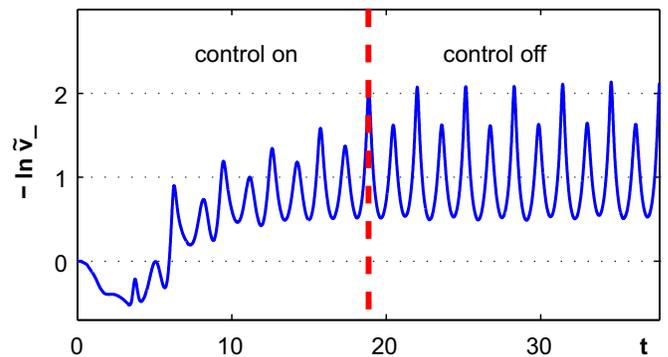}
\caption{\label{fig:hea} Entanglement dynamics for $\beta=0.1$ for a
protocol where the control is switched off at $t_f= 6\pi$. Other
parameters are as in Fig.~\ref{fig:CM}.}
\end{figure}

Our analysis in the accompanying supplemental material~\cite{epaps} reveals
that the final state of the $Q_+$ mode differs very little from a
thermal state for the parameters covered by our numerical
simulations. The condition $\det \gamma < 0$ can thus be re-stated to
a good approximation in the simple form
\begin{equation}\label{eq:bound}
\cosh 2r_- > \coth \beta .
\end{equation}
This indicates that raising temperature does not preclude entanglement
\emph{per se}. However, it gives a lower bound on the squeezing
required to obtain entanglement at a given temperature, which becomes
more and more stringent as temperature is raised. Given an absolute
upper bound of the achievable squeezing $r_-$, Eq. (\ref{eq:bound})
indicates the highest temperature at which the condition $\det
\gamma<0$ can be fulfilled. This result is largely independent of the
dissipative coupling strength; it remains valid as long as the
oscillators are underdamped.

Interestingly, the generated entanglement in the present
scenario even persists once the external control is switched off, as
illustrated in Fig.~\ref{fig:hea} for $\beta=0.1$. This feature
may be attractive for potential applications.

\paragraph{Discussion.-}
We have shown that dynamical symmetry breaking for two-mode Gaussian
parties due to the combined impact of common dissipation and local
parametric control efficiently generates entanglement. This proves
that quantum non-locality can be induced in a minimal model where it
is absent in the thermodynamic state and only appears in
non-equilibrium. Local optimal control allows to achieve substantial
logarithmic negativity in finite time and even at high temperatures
and in a simple, experimentally realizable system.  The protocol may
thus be applicable to various bipartite systems in condensed phase devices. Particular examples include two Cooper pair boxes
spatially well separated in a ''bad'' cavity or  two impurity
fermions  embedded in a
Bose-Einstein condensate. Potentially, NV centers in diamonds
provide a test-bed to dynamically induce entanglement at high
temperatures~\cite{Li2012}. Beyond these direct realizations, there
are also consequences for quantum teleportation to be
explored. Instead of transferring one half of an entangled pair from
A  to B, it may be possible to create a spatially separated
entangled pair \emph{in place} through a dynamical process involving a
(possibly ``dirty'') shared medium or reservoir.


\paragraph{Acknowledgements.} The authors like to thank E. Kajari,
T. H\"aberle and F.~Jelezko for valuable discussions. J.A. thanks for
the kind hospitality of Dartmouth College, Hanover.  Financial support
was provided by Deutsche Forschungsgemeinschaft through AN336/6-1 and
SFB/TRR21.

\bibliographystyle{apsrev4-1}
\bibliography{Literatur_ZHO}
\end{document}